\documentclass[prl,twocolumn,showpacs,floatfix,,amsmath,amssymb,superscriptaddress]{revtex4}
\usepackage{amsfonts}
\usepackage{stmaryrd}
\usepackage{bbm}
\usepackage{mathrsfs}
\usepackage{tipa}
\usepackage{amssymb}
\usepackage{txfonts}
\usepackage{graphicx}
\usepackage{dcolumn}
\usepackage{epstopdf}
\usepackage[colorlinks,linkcolor=blue,urlcolor=blue,citecolor=blue]{hyperref}
\usepackage{multirow}
\usepackage{subfigure}
\usepackage{color}

\begin{document}
\newcommand*{\cm}{cm$^{-1}$\,}
\newcommand*{\Tc}{T$_c$\,}

\title{Optical spectroscopy study of three dimensional Dirac semimetal ZrTe$_5$}
\author{R. Y. Chen}
\affiliation{International Center for Quantum Materials, School of Physics, Peking University, Beijing 100871, China}

\author{S. J. Zhang}
\affiliation{International Center for Quantum Materials, School of
Physics, Peking University, Beijing 100871, China}

\author{J. A. Schneeloch}
\affiliation{Condensed Matter Physics and Materials Science
Department, Brookhaven National Lab, Upton, New York 11973, USA}

\author{C. Zhang}
\affiliation{Condensed Matter Physics and Materials Science
Department, Brookhaven National Lab, Upton, New York 11973, USA}

\author{Q. Li}
\affiliation{Condensed Matter Physics and Materials Science
Department, Brookhaven National Lab, Upton, New York 11973, USA}
\author{G. D. Gu}
\affiliation{Condensed Matter Physics and Materials Science
Department, Brookhaven National Lab, Upton, New York 11973, USA}

\author{N. L. Wang}
\affiliation{International Center for Quantum Materials, School of Physics, Peking University, Beijing 100871, China}
\affiliation{Collaborative Innovation Center of Quantum Matter, Beijing 100871, China}

\begin{abstract}
Three dimensional (3D) topological Dirac materials are under intensive study recently. The layered compound ZrTe$_5$ has been suggested to be one of them by transport and ARPES experiments. Here, we perform infrared reflectivity measurement to investigate the underlying physics of this material. The derived optical conductivity exhibits linear increasing with frequency below normal interband transitions, which provides the first optical spectroscopic proof of a 3D Dirac semimetal. Apart from that, the plasma edge shifts dramatically to lower energy upon temperature cooling, which might be associated with the consequence shrinking of the lattice parameters. In addition, an extremely sharp peak shows up in the frequency dependent optical conductivity, indicating the presence of a Van Hove singularity in the joint density of state.
\end{abstract}

\pacs{71.55.Ak, 78.20.-e,72.15.Eb}

\maketitle

Three dimensional (3D) topological Dirac materials, such as Dirac
semimetals and Weyl semimetals, have attracted tremendous
attention in condensed matter physics and materials science in
recent years. The 3D Dirac semimetals possess 3D Dirac nodes where
the valence and conduction bands touch along certain symmetric
axis. The 3D Dirac node is protected against gap formation by
crystalline symmetry and can be considered as two overlapped Weyl
nodes of opposite chirality
\cite{PhysRevLett.108.140405,PhysRevB.85.155118}. The materials
are expected to host many intriguing novel phenomena and
properties, such as the appearance of Fermi arcs on the surfaces \cite{fermiarchason},
giant diamagnetism \cite{rober1979,PhysRevB.81.195431}, and very large linear
magnetoresistance \cite{PhysRevB.58.2788,PhysRevLett.106.156808}. When time
reversal symmetry is broken, a 3D Dirac semimetal can be driven to
a Weyl semimetal. This has stimulated strong interest to explore
chiral magnetic effect \cite{PhysRevD.78.074033,Son2013b,Hosur2014,Kharzeev2014133,Burkov2014a,Qli,jia2015,chen2015}, \emph{i. e.} the charge-pumping between
Weyl nodes of opposite chiralities under magnetic field, which is
a macroscopic manifestation of the quantum anomaly in relativistic
field theory of chiral fermions.

Only a few materials were predicted to be 3D Dirac semimetals.
Among them, Na$_3$Bi and Cd$_3$As$_2$ are the two best known
compounds \cite{PhysRevB.85.195320,PhysRevB.88.125427}. They were confirmed to be 3D Dirac semimetals by angle-resolved
photoemission spectroscopy (ARPES) experiments
\cite{PhysRevLett.113.027603,Neupane2014,ZKLiu1,ISI:000338482300013}.
Very recently,
another layered compound, ZrTe$_5$, was suggested to be also a 3D
Dirac semimetal based on transport and ARPES experiments \cite{Qli}. ZrTe$_5$ crystallizes in the layered orthorhombic
crystal structure, with prismatic ZrTe$_6$ chains running along
the crystallographic a-axis and linked along the c-axis via zigzag
chains of Te atoms to form two-dimensional (2D) layers. Those layers stack
along the b-axis. Most attractively, the chiral
magnetic effect was clearly observed for the first time on
ZrTe$_5$ through a magneto-transport measurement \cite{Qli}.

 \begin{figure}[b]
  \centering
  \includegraphics[width=7cm]{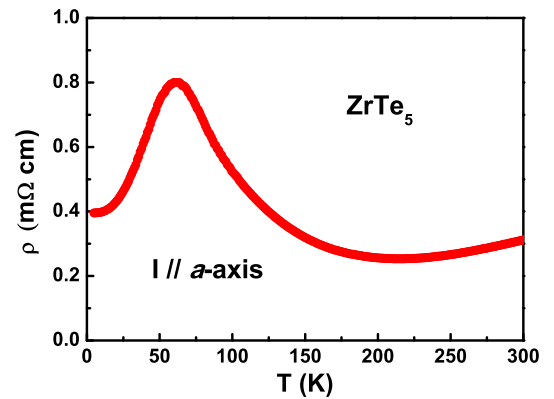}\\
  \caption{Temperature dependent resistivity of ZrTe$_5$ along a-axis.}\label{Fig:res}
\end{figure}

\begin{figure*}
  \centering
  \includegraphics[width=18cm]{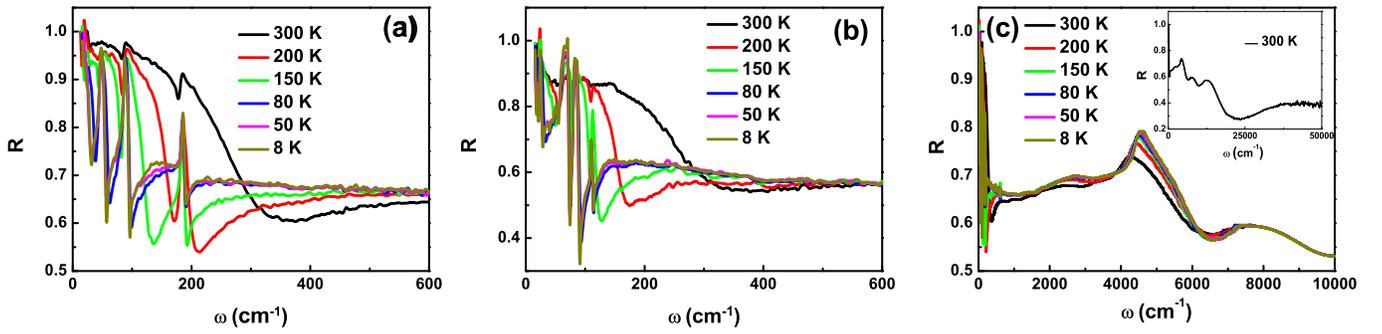}\\
  \caption{Temperature dependent optical reflectivity $R(\omega)$ below 600 \cm along \textit{a}- (panel a) and \textit{c}-axis (panel b). Panel (c) displays $R(\omega)$ below 10000 \cm along \textit{a} axis, while the inset shows the spectrum up to 50000 \cm at room temperature.}\label{Fig:refle}
\end{figure*}

The 3D band dispersions in Dirac
semimetals enable the bulk measurement techniques, \emph{e.g.} the
optical spectroscopy measurement, to provide key information about
the charge properties of the materials. For any electronic
material, such information is of crucial importance for
understanding underlying physics. The situation would be different
from the investigations on 3D topological insulators where the
focus is on the non-trivial surface state. For those materials, the bulk measurement
technique could hardly detect or separate the contribution from
the 2D Dirac fermions of topological surface states, as a result,
the experimental probes were largely limited to surface sensitive
measurements like ARPES or scanning tunneling microscopy (STM).

Although a 3D Dirac semimetal could be viewed as a bulk analogue of
2D graphene in terms of band dispersions, a key difference
exists in charge dynamical properties between 2D and 3D Dirac
fermions. For 2D Dirac fermions like in graphene, the real part of
conductivity is well known to be a frequency independent quantity \cite{PhysRevLett.101.196405},
while for 3D Dirac fermions with linear band dispersions in all
three momentum directions, the real part of conductivity grows
linearly with the frequency \cite{PavanHosur,Ashby2014}. To our knowledge, no optical
spectroscopic measurement has been reported on 3D Dirac
semimetals. In this work we present temperature and frequency
dependent optical spectroscopy study on ZrTe$_5$. Our study
reveals several peculiar features. The density of free carriers keeps dropping with temperature decreasing and an unusual sharp peak shows up in the reflectivity and conductivity spectra in the mid-infrared region. Of most importance, the linear rising of the optical conductivity is observed in a relatively large energy scale, just as expected for a 3D Dirac semimetal.

Single crystals ZrTe$_5$ were grown by using Te flux method.  100
grams high purity 7N (99.99999\%) Te and 6N (99.9999\%) Zr were
loaded into 20 mm diameter double-walled quartz ampoules with 200
mm length and sealed under vacuum. The composition of the Zr-Te
solution for the ZrTe$_5$ crystal growth is Zr$_{0.0025}$Te$_{0.9975}$, and the
largest ZrTe$_5$ single crystals obtained are around $\sim$ 1 $\times$ 1 $\times$ 20 mm$^3$.
Detailed growth procedure was described elsewhere \cite{Qli}.

Figure \ref{Fig:res} displays the dc resistivity in the chain direction as a function of temperature of ZrTe$_5$. The resistivity curve shows rather
unusual features. With decreasing temperature from 300 K, $\rho(T)$ decreases slightly first, then increases below
roughly 200 K and reaches a peak near 60 K. The overall small values of $\rho(T)$ signal the (semi)metallic nature of the compound,
which are in consistent with the optical measurements presented below. In earlier reports, the resistivity peak located near
150 - 170 K \cite{PhysRevB.24.2935,PhysRevB.31.7617}. The much lower peak temperature could be attributed to a much lower defects in the present sample.
The unusual features are likely associated with its characteristic electronic structure, which could be sensitive to the
temperature-induced structural change. Earlier band structure calculations suggested presence of extremely small
and light ellipsoidal Fermi surfaces, centered at the center
($\Gamma$ point) of the bulk Brillouin zone (BBZ) \cite{PhysRevB.26.687}, and
quantum oscillation measurements indicated three tiny but finite Fermi surfaces \cite{PhysRevB.31.7617}, with
the effective mass in the chain direction (m$^*$a$\simeq$0.03m$_e$) being comparable to that in a
prototypical 3D Dirac semimetal, Cd$_3$As$_2$ \cite{PhysRevLett.113.246402}.
More recent \emph{ab initio}
calculations indicated that ZrTe$_5$ compound locates close to the phase boundary between weak and strong topological insulators \cite{PhysRevX.4.011002}. Nevertheless, very recent transport and ARPES experiments identify it to be a 3D Dirac semimetal \cite{Qli}.

The polarized optical reflectance measurements with \textbf{E}//\textit{a}-axis and \textbf{E}//\textit{c}-axis were performed respectively on a combination of Bruke 113V, 80V and grating-type spectrometer in the frequency range from 30-50 000 \cm.
The reflectivity along \textit{a}- and \textit{c}-axis in the far-infrared region are displayed in Fig. \ref{Fig:refle} (a) and (b), respectively. The \textit{a}-axis reflectivity over broad energy scales is shown in Fig. \ref{Fig:refle} (c). There are several prominent phonon absorption features lying below 200 \cm, which get more pronounced upon cooling. At room temperature, the reflectivity $R(\omega)$ shows a well-defined plasma edge below 350 \cm and approaches unity at low frequency. This indicates clearly a metallic response, which is consistent with the results of transport measurements. The edge frequency, being usually referred to as "screened" plasma frequency, is related to the density $n$ and effective mass $m^*$ of free carriers by $\omega_p^{'2}\propto n/m^*$. The unusual low value of this frequency suggests a very small $n$. As the temperature decreases, the edge shifts to lower frequency, indicating a further reduction of the number of free carriers. Below 80 K, the extremely strong phonon signals tend to blur the plasma edge. In the meantime, the carrier scattering rate is also reduced as the reflectance edge gets steeper. The combination of the two effects is likely the driving force of the broad peak appeared at 60 K in $\rho(T)$. Furthermore, the rapid reduction of the plasma frequency yields optical evidence that the compound tend to approach the semimetal state at very low temperature.

R($\omega$) along the \textit{c} axis shows similar behaviors as along the \textit{a} axis. The plasma edge almost lies at the same position as along the \textit{a} axis at different temperatures. However, the overall reflectivity is a little bit lower along the \textit{c} axis, which is absolutely reasonable because the conductivity along this direction is supposed to be lower. Moreover, the phonon signals shown in Figure \ref{Fig:refle} (a) and (b) are quite distinct, revealing different oscillating modes. In the following paragraphes, we will only discuss the characteristic features along the \textit{a} axis.

Basically, the spectra at higher energies are dominated by interband transitions, as shown in Fig. \ref{Fig:refle} (c). The peaks centered around 2700 \cm, 4500 \cm and 7800 \cm represent for different transitions. Remarkably, the interband transitions exhibit temperature dependence. This is seen more clearly for the second peak. As temperature decreases, the peak frequency moves to higher energy and its spectral weight is enhanced simultaneously. The enhancement is linked to the reduction of Drude spectral weight. As we shall explain below, it may be caused by the shift of chemical potential associated with temperature-induced subtle crystal structural change.

\begin{figure}
  \centering
  \includegraphics[width=7.5cm]{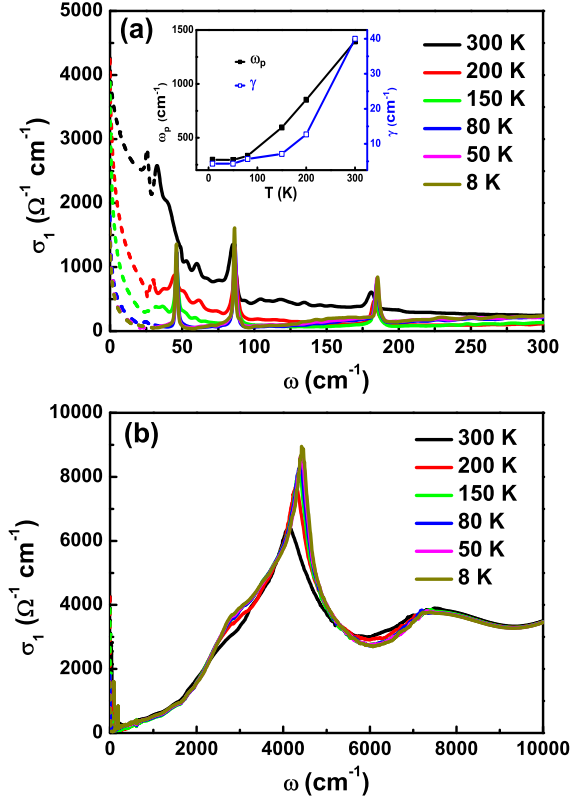}\\
  \caption{The optical conductivity at different temperatures up to (a) 300 \cm and (b) 10 000 \cm. The short dashed lines are low frequency extrapolations by the Hagen-Rubens relation. The inset of the upper panel shows the temperature dependent evolution of the plasma frequency $\omega_p$ and scattering rate of free carriers $\gamma$ by black solid squares and blue open squares respectively.   }\label{Fig:cond}
\end{figure}

The real part of optical conductivities $\sigma_1$($\omega$) are derived from R($\omega$) though Kramers-Kronig relation, as displayed in Figure \ref{Fig:cond}. The Hagen-Rubens relation was used for the low frequency extrapolation, and the x-ray atomic scattering functions were used in the high frequency extrapolation \cite{Tanner}. Figure \ref{Fig:cond} (a) shows $\sigma_1$($\omega$) in the low frequency. Although the screened plasma edge can be clearly observed in R($\omega$), the Drude component in $\sigma_1$($\omega$) appears mainly in the extrapolated region. Apparently, the Drude spectral weight reduces significantly at low temperature, being consistent with the dramatic shift of plasma edge towards low frequency in R($\omega$).
In addition, three sharp phonon peaks are resolved at 46, 86 and 185 \cm, respectively. When the temperature decreases, the phonon frequencies shift slightly towards higher energies.

The free carrier contributions could be quantitatively estimated from the application of Drude model, which has a peak centered at zero frequency and the peak width is just the carrier scattering rate $\gamma$. The Drude spectral weight, which gives the plasma frequency $\omega_p=\sqrt{4\pi n/m^*}$, could be alternatively obtained from integrating the low frequency spectral weight by $\omega_p^2=8\int_0^{\omega_c}\sigma_1d\omega$, where $\omega_c$ is the cut off frequency. The integration should sum up all the free carrier contributions, without the influence of any interband transitions. We chose the frequency where $\sigma_1(\omega)$ reaches a minimum as $\omega_c$,
which was identified to be 30, 30, 70, 130, 220 and 330 \cm in sequence corresponding to increasing temperatures. The yielded temperature dependent scattering rate and plasma frequency are plotted in the inset of \ref{Fig:cond} (a). The scattering rate decreases from 40 \cm at 300 K to 4 \cm at 8K. Meanwhile, $\omega_p$ drops monotonically upon cooling from 1400 \cm to 300 \cm. Assuming the effective mass of free carriers remains unchanged, then 95\% of the population is reduced. The dramatic reduction of free carriers could hardly been explained by reduced thermal excitations. Here, we propose that this peculiar phenomenon is related to the chemical potential, which moves closer to the Dirac node as temperature decreasing, along with the lattice shrinking. Theoretical calculation suggested that the electronic properties of ZrTe$_5$ are extraordinarily sensitive to the variation of the lattice parameters \cite{PhysRevX.4.011002}.

Figure \ref{Fig:cond} (b) presents the optical conductivity in the energy scale up to 10 000 \cm. Corresponding to R($\omega$), an exceptional sharp peak shows up at around 4000 \cm, along with a shoulder-like feature at around 2800 \cm. When temperature decreases, the central frequency of the sharp peak shifts to higher energy and the half with of it gets smaller. At the lowest temperature, it even becomes quasi-divergence. This feature has never been observed in any other 3D materials. We noticed that the reflectance of single-walled carbon nanotubes, whose joint density of states (JDOS) are constituted by individual Van Hove singularities, was reported to show geometry-dependent resonant peaks \cite{nanotube}, which are identified to stem from interband transitions between these singularities. Consequently, it is most likely that the quasi-divergent peak exhibited in $\sigma_1(\omega)$ also requires for a Van Hove singularity in the JDOS of ZrTe$_5$, which is rarely seen in a 3D bulk material. Nevertheless, the ZrTe$_5$ single crystals are easily cleaved both along the \textit{b} and \textit{c} axis, forming a needle-like compound. The lattice parameters of a=3.9876 $\AA$, b=14.505 $\AA$, c=13.2727 $\AA$ also indicate a quasi-one dimensional property, which tends to strengthen the effect of a Van Hove singularity.

\begin{figure}[t]
  \centering
  \includegraphics[width=7.5cm]{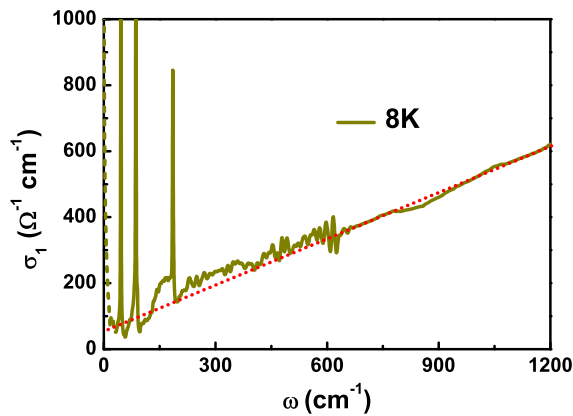}\\
  \caption{The optical conductivity at 8 K below 1200 \cm. The red dotted line are the linear fitting of $\sigma_1(\omega)$ }\label{Fig:line}
\end{figure}

Usually Lorentz model is employed to describe the interband transition. However, for present compound, the conductivity data above the Drude component could not be reproduced by the Lorentz model due to a linear increase of conductivity with frequency. To make it clearly to see, the optical conductivity at 8 K is displayed in Fig. \ref{Fig:line}, where the linear increasing is fitted by a red dotted line. Ba\'{a}si and Virosztek showed that, for a noninteracting electron system consisting of two symmetric energy bands touching each other at the Fermi level whose Hamiltonian depends on the magnitude of the momentum through a power-law behavior, the optical conductivity of the interband contribution has a power-law frequency dependence with $\sigma_1(\omega)\propto (\frac{\hbar\omega}{2})^{\frac{d-2}{z}}$, where d is the dimension of the system and z represents for the power-law term of the band dispersion \cite{PhysRevB.87.125425}. For example, the two dimensional graphene with linear dispersion of $\varepsilon(k)\propto |k|$ (that is, $d$=2 and $z$=1) gives rise to a frequency independent conductivity, which has been confirmed by optical experiments \cite{PhysRevLett.101.196405}. For the case of 3D compound ZrTe$_5$, the linear rising $\sigma_1(\omega)$ demands for $\frac{d-2}{z}=1$. By applying $d$=3, a linear dispersion $z$=1 is obtained, in good agreement with a Dirac semimetallic behavior.

As a matter of fact, the linearly growing optical conductivity has also been observed in a Rh-doped Nd$_2$Ir$_2$O$_7$ below 50 meV \cite{P227}. Since pyrochlore iridates were suggested to host Weyl semimetals \cite{Wan2011b}, the observation has been taken as the evidence of a Weyl semimetal even though the linear increasing only persistd in a very small energy scale. Significantly, several quasicrystals (such as Al$_{63.5}$Cu$_{24.5}$Fe$_{12}$, Al$_{75.5}$Mn$_{20.5}$Si$_{10.1}$) has shown similar behaviors as well \cite{quasicrystal}, all of which are lack of inversion symmetries and were suggested to be candidates of Weyl semimetals. Furthermore, 3D massless electrons with huge Fermi velocity are certified by linear rising optical conductivity in the zinc-blende crystal Hg$_{1-x}$Cd$_x$Te, with a proper doping concentration close to $x$=0.17 \cite{Orlita2014}.
What we observed here, however, is the first bulk evidence of a time reversal and space inversion symmetry protected 3D Dirac semimetal in the as-grown single crystal.

The linear increase of optical conductivity is an unparalleled property of bulk 3D Dirac band dispersions. Theoretical studies show that the optical conductivity contributed by per Weyl node (lifting the spin degeneracy) is \cite{PavanHosur,Ashby2014} $\sigma_1(\omega)$=$\frac{e^2}{12h}\frac{\omega}{v_F}$=$G_0$$\frac{\omega}{24v_F}$, where $G_0=2e^2/h$=7.727$\times$10$^{-5} \Omega^{-1}$ is the quantum conductance. For ZrTe$_5$, band structural calculations indicates only one possible Dirac node near $\Gamma$ point per ZrTe$_5$ layer \cite{PhysRevX.4.011002}. Considering that each Dirac node has a spin degeneracy of 2, we can get the Fermi velocity $v_F=2.59\times$10$^6$ cm/s in terms of the slope of the conductivity shown in Fig. \ref{Fig:line}. This value is significantly smaller than the Fermi velocity determined by ARPES experiment along \textit{a}- or \textit{c}-axis (smaller but close to 10$^8$ cm/s). We remark that, for 3D Dirac semimetals, although the linear dispersions exist for all three momentum directions, their slopes or Fermi velocities are in fact different. ZrTe$_5$ has a layered structure, the Fermi velocity in the direction perpendicular to the layers (along \textit{b}-axis) must be much smaller than the other two directions, unfortunately presently available measurement has not yielded such information. The Fermi velocity determined from optical conductivity for a 3D Dirac semimetal must contain contributions from all momentum directions.

For a rigorous Dirac semiemtal, the Fermi energy lies right across the Dirac points and the Fermi surfaces would shrink to isolated points in the BZ. As a result, there would be no free carriers. The optical conductivity of ZrTe$_5$ shows obvious Drude component at high temperatures, implying that the Fermi energy stays away from the Dirac point. With temperature decreasing, the density of free carriers decreases monotonically and almost all of them are lost at the lowest temperature, which is consistent with the approaching of the Fermi level to the Dirac point. ARPES measurement has produced a very similar scenario. It is also worth noting that the frequency where the conductivity deviates from linear dependence can not be treated as the cutoff energy of linear band dispersion, as $\sigma_1(\omega)$ is dramatically modified by other interband transitions at higher energies.

In conclusion, we have studied the optical conductivity of the 3D topological Dirac semimetal material ZeTe$_5$. The number of free carriers are proved to be much smaller than conventional metals, manifesting a semimetallic behavior. Furthermore, it decreases as temperature decreasing, probably caused by the modification of chemical potential, which might be sensitive to the lattice parameter variation. An unexpected quasi-divergent peak is also observed in the optical conductivity, which is possibly associated with a Van Hove singularity in the JDOS. Most importantly, the optical conductivity grows linearly with frequency below 1200 \cm, presenting the first optical spectroscopic evidence of a 3D Dirac semimetal.

\begin{center}
\small{\textbf{ACKNOWLEDGMENTS}}
\end{center}

We acknowledge very helpful discussions with H. M. Weng, F. Wang, X. C. Xie, Z. Fang,
X. Dai, H. W. Liu. This work was supported by the National
Science Foundation of China (11120101003, 11327806), and the 973
project of the Ministry of Science and Technology of China
(2011CB921701, 2012CB821403). Work at Brookhaven is supported by the Office of Basic Energy Sciences, Division of Materials Sciences and Engineering, U.S. Department of Energy under Contract No. DE-SC00112704

\bibliographystyle{apsrev4-1}
  \bibliography{ZrTe5}

\end{document}